\documentclass{article}
\usepackage{waspaa19,amsmath,url,times}

\usepackage[keeplastbox]{flushend}  
\usepackage{blindtext, graphicx}
\usepackage{bm}
\usepackage{booktabs}
\usepackage{subfig}
\usepackage[font=footnotesize]{caption}
\usepackage{algorithm2e}
\usepackage{siunitx}

\usepackage{textcomp}
\usepackage[absolute,showboxes]{textpos}

\TPMargin{5pt}

\newcommand{\copyrightstatement}{
	\begin{textblock}{0.84}(0.08,0.93)    
		\noindent
		\footnotesize
		\textcopyright~2019~IEEE.  Personal  use  of  this  material  is  permitted.
		Permission from IEEE must be obtained for all other uses, in any current or
		future media, including reprinting/republishing this material for advertising
		or promotional purposes, creating new collective works, for resale or
		redistribution to servers or lists, or reuse of any copyrighted component of
		this work in other works.
	\end{textblock}
}

\usepackage{localmath}

\title{INDEPENDENT VECTOR ANALYSIS WITH MORE MICROPHONES THAN SOURCES}
\name{Robin Scheibler and Nobutaka Ono\thanks{This work was supported by a JSPS post-doctoral fellowship and grant-in-aid (no. 17F17049), and the SECOM Science and Technology Foundation. The research presented in this paper is reproducible. Code and data are available at \protect\url{https://github.com/onolab-tmu/overiva}.}}
\address{Tokyo Metropolitan University, Tokyo, Japan}       

\begin{document}

\ninept

\copyrightstatement

\maketitle


\begin{abstract}
  We extend frequency-domain blind source separation based on independent vector analysis to the case where there are more microphones than sources.
  The signal is modelled as non-Gaussian sources in a Gaussian background.
  The proposed algorithm is based on a parametrization of the demixing matrix decreasing the number of parameters to estimate.
  Furthermore, orthogonal constraints between the signal and background subspaces are imposed to regularize the separation.
  The problem can then be posed as a constrained likelihood maximization.
  We propose efficient alternating updates guaranteed to converge to a stationary point of the cost function.
  The performance of the algorithm is assessed on simulated signals.
  We find that the separation performance is on par with that of the conventional determined algorithm at a fraction of the computational cost.
\end{abstract}


\section{Introduction}

We address the problem of blindly separating $K$ sound sources recorded with $M$ microphones when $K < M$.
By far the most popular technique for blind source separation (BSS) is independent component analysis (ICA) which only requires statistical independence of the sources \cite{Comon:1994kr}.
A convolutive sound mixture is written
\begin{equation}
	\hat{x}_m[t] = \sum_{k=1}^K (\hat{a}_{mk} \star \hat{s}_k)[t],
\end{equation}
where $\hat{x}_m[t]$ is the $m$-th microphone signal, $\hat{s}_k[t]$ is the $k$-th source signal, and $\hat{a}_{mk}[t]$ is the impulse response between the two. The operator $\star$ denotes convolution.
In the time-frequency domain, convolution becomes frequency-wise multiplication and we have
\begin{equation}
  x_{mfn} = \sum_{k=1}^K a_{mkf} s_{kfn},
\end{equation}
where $x_{mfn}$ and $s_{kfn}$ are the short-time Fourier transforms (STFT) \cite{Allen:1977in} of $\hat{x}_m[t]$ and $\hat{s}_k[t]$, respectively, and $a_{mk}[f]$ is the discrete Fourier transform of $\hat{a}_{mk}[t]$.
Finally, $f=1,\ldots,F$ and $n=1,\ldots,N$ are the discrete frequency bin and frame indices, respectively.
This is an approximation valid when the Fourier transform is sufficiently longer than the impulse response.
In this form, the separation problem can be solved by applying ICA to every frequency sub-band independently~\cite{Smaragdis:1998kl}.
Unfortunately, the assignment of output signals to sources in each of the sub-bands is unknown and the correct permutation must be recovered.
Clustering is a popular solution for permutation alignment~\cite{Sawada:fk}.
Nevertheless, this extra step is notoriously hard to get right and avoiding it is desirable.
Independent vector analysis (IVA) does just that by considering the problem as joint separation over frequencies~\cite{Hiroe:2006ib,Kim:2006ex}.
The computationally efficient, hyperparameter-free, method for ICA and IVA known as iterative projection \cite{Ono:2010hh,Ono:2011tn,Ono:2012bh} forms the basis of our work.

Both for ICA and IVA, the determined case, i.e., $K=M$, is the most straightforward.
It allows to do a change of variables and directly maximize the likelihood of the separated signals.
In practice, however, using extra microphones adds robustness and increases performance.
This is the so-called \textit{overdetermined} case with $K < M$.
Unfortunately, the aforementioned change of variables cannot be done anymore.
A straightforward solution to this problem is to run the algorithm for $M$ sources, and retain the $K$ outputs with the largest power.
Alternatives to power-based selection exist, for example~\cite{Kitamura:2015fi}.
Due to the large number of parameters, $\calO(M^2)$, to estimate, such approaches come with a high computational cost.
Ideally, we want to estimate no more than $\calO(KM)$ parameters.

Several methods with better complexities have been proposed.
These methods fall broadly in two categories.
First, some methods not based on the aforementioned change of variables can directly tackle the overdetermined case~\cite{Murata:1998ul,Amari:1999jy}, but some require regularization~\cite{Nishikawa:2004ug}.
Second, methods that first reduce the number of channels to $K$ and then apply a determined separation algorithm.
This is done for example by selecting the best $K$ channels~\cite{Nishikawa:2004bn,Osterwise:2014fe}, or by principal component analysis (PCA)~\cite{Osterwise:2014fe,Joho:2000tt,Lee:2007bw}.
Nevertheless, these methods inherently risk removing some target signal upfront, irremediably degrading performance.
Anecdotally, a few methods have been proposed for instantaneous mixtures~\cite{Fu:jx,Souden:2006wf}, and in the time-domain~\cite{Diamantaras:ib}.
All the above methods are single mixture methods that require permutation alignment.
Few techniques have been proposed for overdetermined IVA.
The single source case, i.e., $K=1$, known as independent vector extraction (IVE), has been tackled with a gradient ascent method~\cite{Koldovsky:fn}.

We propose \textit{OverIVA}, an algorithm to perform IVA with $K < M$.
The proposed algorithm is hyperparameter-free, guaranteed to converge, and only requires the estimation of $\calO(KM)$ parameters.
We derive two variants based on the Laplace and time-varying Gaussian source distributions.
The resulting algorithms can be seen as extensions of IVE~\cite{Koldovsky:fn} to more than one source, and with the fast converging updates of AuxIVA~\cite{Ono:2011tn}.
Numerical experiments reveal its separation performance to be comparable to that of full $M$-channels IVA at a fraction $K/M$ of the computational cost.
We also find that adding extra microphones fails to improve the performance when using PCA as a pre-processing in diffuse noise.

The rest of this paper is organized as follows. \sref{model} describes the hypotheses and signal model.
In \sref{algo}, we derive the proposed algorithm.
The numerical experiments are discussed in \sref{perfeval}. \sref{conclusion} concludes.

\section{Model}
\seclabel{model}

The microphone signals $\vx_{fn} = [x_{1fn},\ldots,x_{Mfn}]^\top \in \C^M$ at frequency $f$ and time $n$ is modelled as
\begin{equation}
\vx_{fn} = \mA_f \vs_{fn} + \mPsi_f \vz_{fn},
\end{equation}
where $\vs_{fn} = [s_{1fn},\ldots,s_{Kfn}]^\top \in \C^K$ contains the source signals, $\vz_{fn} \in \C^{M-K}$ is a vector of noise, and $\mA_f \in \C^{M\times K}$ and $\mPsi_f \in \C^{M \times M-K}$ are the respective mixing matrices.
Our objective is to estimate the \textit{demixing matrix} $\wh{\mW}_f \in \C^{M\times M}$ such that the source vector $\vs_{fn}$ is
recovered from the measurements
\begin{equation}
    \begin{bmatrix} \vs_{fn} \\ \mPhi_f \vz_{fn} \end{bmatrix} = \wh{\mW}_f \vx_{fn}.
\end{equation}
The matrix $\mPhi_f$ is an arbitrary invertible linear transformation reflecting that we do not aim at separating the noise components.
Indeed, we may even choose $\mPhi_f$ to simplify the task at hand.
Namely, we choose it so that
\begin{equation}
  \wh{\mW}_f = \begin{bmatrix} \mW_f \\ \mU_f \end{bmatrix}\quad \text{with}\quad 
  \begin{array}{r@{\;}l}
    \mW_f & = \begin{bmatrix} \vw_{1f} & \cdots & \vw_{Kf} \end{bmatrix}^H\in \R^{K\times M} \smallskip, \\
    \mU_f & = \begin{bmatrix} \mJ_f & -\mI_{M-K} \end{bmatrix} \in \R^{M-K\times M},
  \end{array}
\end{equation}
with $\mJ_f\in\C^{M-K\times K}$.
With a slight abuse of notation, we let $\vz_{fn} = \mU_f \vx_{fn}$.
Following blind source separation principles, we will assume that the target sources have some non-Gaussian distribution.
On the other hand, because we do not want to separate the noise components, they are likely to stay mixed and thus their distribution can be assumed close to Gaussian.
However, the Gaussianity of the background by itself will turn out to be ineffective at separating the foreground components.
We thus rely on orthogonal constraints to further help separation \cite{Cardoso:1994wj,Koldovsky:fn}.
We formalize this intuition with the following hypothesis.
\begin{enumerate}
  \item The separated sources are statistically independent
    \begin{equation}
      \vs_{kn} \perp \vs_{k^\prime n^\prime},\ \forall k \neq k^\prime, n, n^\prime
    \end{equation}
    where we use the notation $\vs_{kn}\in\C^F$ to mean the vector of frequency components of the $k$-th source vector at frame $n$.
    In addition, the separated sources have a \textit{time-varying} circular Gaussian distribution (or Laplace, see \sref{algo_laplace})
    \begin{equation}
      p_{\vs} (\vs_{kn}) = \frac{1}{\pi^F r_{kn}^F} e^{-\frac{\|\vs_{kn}\|^2}{r_{kn}}},
      \elabel{dist_source}
    \end{equation}
    where $r_{kn}$ is the variance of source $k$ at time $n$.
    
  \item The separated background noise vectors have a \textit{time-invariant} complex Gaussian
    distribution across microphones
    \begin{equation}
      p_{\vz_f}(\vz_{fn}) = \frac{1}{\pi^{M-K} |det(\mR_{f})|} e^{-\vz_{fn}^H (\mR_{f})^{-1} \vz_{fn}}
      \elabel{dist_noise}
    \end{equation}
    where $\mR_{f}$ is the (unknown) spatial covariance matrix of the noise (after separation).
    Moreover, the separated background noise is statistically independent across frequencies.

  \item The sources and background span orthogonal subspaces after separation, namely,
    \begin{equation}
      \vzero = \frac{1}{N} \mY_f \mZ_f^H = \mW_f \mC_f \mU_f^H,\  \text{with}\  \mC_f = \frac{1}{N} \mX_f \mX_f^H,
      \elabel{orth_const}
    \end{equation}
    where $\mX_f = [\vx_{f,1}, \ldots, \vx_{f,N}]$, $\mY_f = \mW_f \mX_f$, and $\mZ_f = \mU_f \mX_f$.
    The matrix $\mC_f$ is the covariance of the input signal.
\end{enumerate}
Based on these hypothesis, we can write explicitly the likelihood function of the data and find the demixing matrices maximizing it.
A few points are in order. We assume the covariance matrix of the noise is rank $M-K$. In practice, this means that we will not be able to remove noise that has the same steering vector as one of the sources.
Independence of noise across frequencies is a simplifying assumption and is typically not fulfilled.
We confirm in the experiment of \sref{perfeval} that this does not seem to be a problem.
One can also wonder how the algorithm can tell apart sources from noise.
While we do not offer a precise analysis, we conjecture that the $K$ strongest sources have a very non-Gaussian distribution.
On the contrary, the mix of the noise and remaining weaker sources will have a distribution closer to Gaussian.
As such, we expect the maximum likelihood to choose the strongest sources automatically.

\section{Algorithm}
\seclabel{algo}

By using \eref{dist_source} and \eref{dist_noise}, and omitting all constants, we can write the negative log-likelihood of the observed data
\begin{multline}
  \calJ  = -2N\sum_{f} \log |\det(\wh{\mW}_f)|
  + \sum_{kn} \left(F\log r_{kn} + \frac{\| \vs_{kn} \|^2}{r_{kn}}\right) \\
  + \sum_{fn} \left(\log |\det(\mR_{f})| + \vz_{fn}^H (\mR_{f})^{-1} \vz_{fn} \right).
  \elabel{cost_fun}
\end{multline}
where $\| \vs_{kn} \|^2 = \sum_{f}|\vw_{kf}^H \vx_{fn}|^2$.
The first term is due to the change of variables.
First, one can show that the gradient of \eref{cost_fun} with respect to $\mR_f$ is zero when
  $\mR_f = \mU_f \mC_f \mU_f^H$.
Furthermore, for this choice of $\mR_f$, regardless of the choice of $\mU_f$, we have
\begin{equation}
  \sum_n \vz_{fn}^H \mR_f^{-1} \vz_{fn} = \tr\left( \mR_f^{-1} \mZ_f\mZ_f^H \right) = N(M-K).
\end{equation}
As a consequence, once $\mR_f$ has been fixed, the background part of the cost function can be ignored for the estimation of $\wh{\mW}_f$.

The minimization of \eref{cost_fun} with respect to $\mW_f$ can be carried out as in AuxIVA~\cite{Ono:2011tn} via the iterative projection method.
Because direct minimization for $\mW_f$ is difficult, this method minimizes \eref{cost_fun} alternatively with respect to $\vw_{kf}$, $k=1,\ldots,K$.
\begin{align}
  \begin{array}{@{}r@{$\;$}lr@{$\;$}l@{}}
  r_{kn} & \gets \frac{1}{F} \sum_f | \vw_{kf}^H \vx_{fn} |^2, &
  \mV_{kf} & \gets \frac{1}{N}\sum_n \frac{1}{r_{kn}} \vx_{fn} \vx_{fn}^H, \smallskip \\
  \vw_{kf} & \gets \left( \wh{\mW}_f \mV_{kf} \right)^{-1} \ve_k, &
  \vw_{kf} & \gets \frac{\vw_{kf}}{\left(\vw_{kf}^H \mV_{kf} \vw_{kf}\right)^{-\frac{1}{2}}}.
	\end{array}
	\elabel{V_up}
\end{align}
Once these updates have been applied, we must modify the lower part of the demixing matrix, i.e. $\mJ_f$, so that the noise subspace stays orthogonal.
For fixed $\mW_f$, we can solve \eref{orth_const} for $\mJ_f$ and obtain
\begin{equation}
  \mJ_f = \left( \mE_2 \mC_f \mW_f^H \right) \left( \mE_1 \mC_f \mW_f^H \right)^{-1},
  \elabel{J_up}
\end{equation}
where $\mE_1 = [\mI_K\ \vzero_{K\times M-K}]$ and $\mE_2 = [\vzero_{M-K\times K}\ \mI_{M-K}]$.

The final algorithm applying updates to $\mW_f$ and $\mJ_f$ alternatively is detailed in \algref{oiva}.
Each of the updates from \eref{V_up} and \eref{J_up} set the gradient of the cost function to zero with respect to the parameter optimized.
Thus, the value of the cost function is non-increasing under these updates.
While convergence to a global minimum is not guaranteed, convergence to a stationary point is.
Concerning the initial value of $\mW_f$, we find that a rectangular identity matrix is satisfactory.

\begin{algorithm}[t]
\SetKwInOut{Input}{Input}\SetKwInOut{Output}{Output}
\Input{Microphones signals $\{ \vx_{fn} \}$, \# sources $K$}
\Output{Separated signals $\{ \vs_{fn} \}$ }
\DontPrintSemicolon
$\vs_{fn} \gets \vx_{fn},\ \forall f,n$\;
$\mW_f \gets [ \mI_M\ \vzero_{K\times M-K} ],\ \forall f$\;
$\mJ_f \gets \vzero_{M-K\times K},\ \forall f$\;
\For{loop $\leftarrow 1$ \KwTo $\text{max. iterations}$}{
  \For{$k \leftarrow 1$ \KwTo $\text{K}$}{
    $r_{kn} \gets \frac{1}{F} \sum_f |s_{kfn}|^2,\ \forall n$\;
    \For{$f \gets 1$ \KwTo $F$}{
      $\mV_{kf} \gets \frac{1}{N} \sum_n \frac{1}{r_{kn}} \vx_{fn} \vx_{fn}^H$\;
      $\vw_{kf} \gets (\wh{\mW}_f \mV_{kf})^{-1} \ve_k$\;
      $\vw_{kf} \gets \vw_{kf} \left(\vw_{kf}^H \mV_{kf} \vw_{kf}\right)^{-\frac{1}{2}}$\;
      $s_{kfn} \gets \vw_{kf}^H \vx_{fn},\ \forall n$\;
      $\mJ_f \gets \left( \mE_2 \mC_f \mW_f^H \right) \left( \mE_1 \mC_f \mW_f^H \right)^{-1}$\;
    }
  }
}
\caption{OverIVA}
\label{alg:oiva}
\end{algorithm}

\subsection{Laplace overdetermined IVA}
\seclabel{algo_laplace}

The algorithm presented so far assumes a time-varying Gaussian distribution of source vectors.
It is possible to change the model to a \textit{time-invariant} circular Laplace distribution as in AuxIVA~\cite{Ono:2011tn}.
Under this new source model, the cost function becomes
\begin{equation}
  \calL  = -2N\sum_{f} \log |\det(\wh{\mW}_f)| + \sum_{kn} \| \vs_{kn} \|_2 + \sum_{fn} \log p_{\vz_f}(\vz_{fn}).
  \nonumber
\end{equation}
Ignoring constants, one can show that this new cost function is majorized by \eref{cost_fun} for the specific choice~\cite{Ono:2011tn}
\begin{equation}
  r_{kn} = 2 \sqrt{\sum\nolimits_f | \vw_{kf}^H \vx_{fn} |^2}.
\end{equation}
In this case, OverIVA becomes an auxiliary function based optimization procedure that is still guaranteed to converge to a stationary point.

\subsection{Computational Complexity}
\seclabel{comp}

When the number of time frames $N$ is larger than the number of microphones $M$, the runtime is dominated by the computation of the weighted covariance matrix $\mV_{kf}$.
The computational complexity in that case is $\calO(K F M^2 N)$. When the number of microphones is larger, the bottleneck is the matrix inversion with complexity $\calO(K F M^3)$.
The total complexity of the algorithm is thus
\begin{equation}
  \calC_{\text{OverIVA}} = \calO(KF M^2 \max\{M, N\}).
\end{equation}
The leading $K$ comes from the number of demixing filters (one per source), and $F$ is the number of frequency bins.
In contrast, conventional AuxIVA needs to update all $M$ demixing filters, which leads to complexity
\begin{equation}
  \calC_{\text{AuxIVA}} = \calO(F M^3 \max\{M, N\}).
\end{equation}
The overall complexity is thus reduced by a factor $K/M$.
This is significant in many practical cases as the number of target sources is rarely larger than four, and the number of microphones can easily be over ten for larger arrays.

\begin{figure}
\begin{center}
  \includegraphics[width=0.6\linewidth]{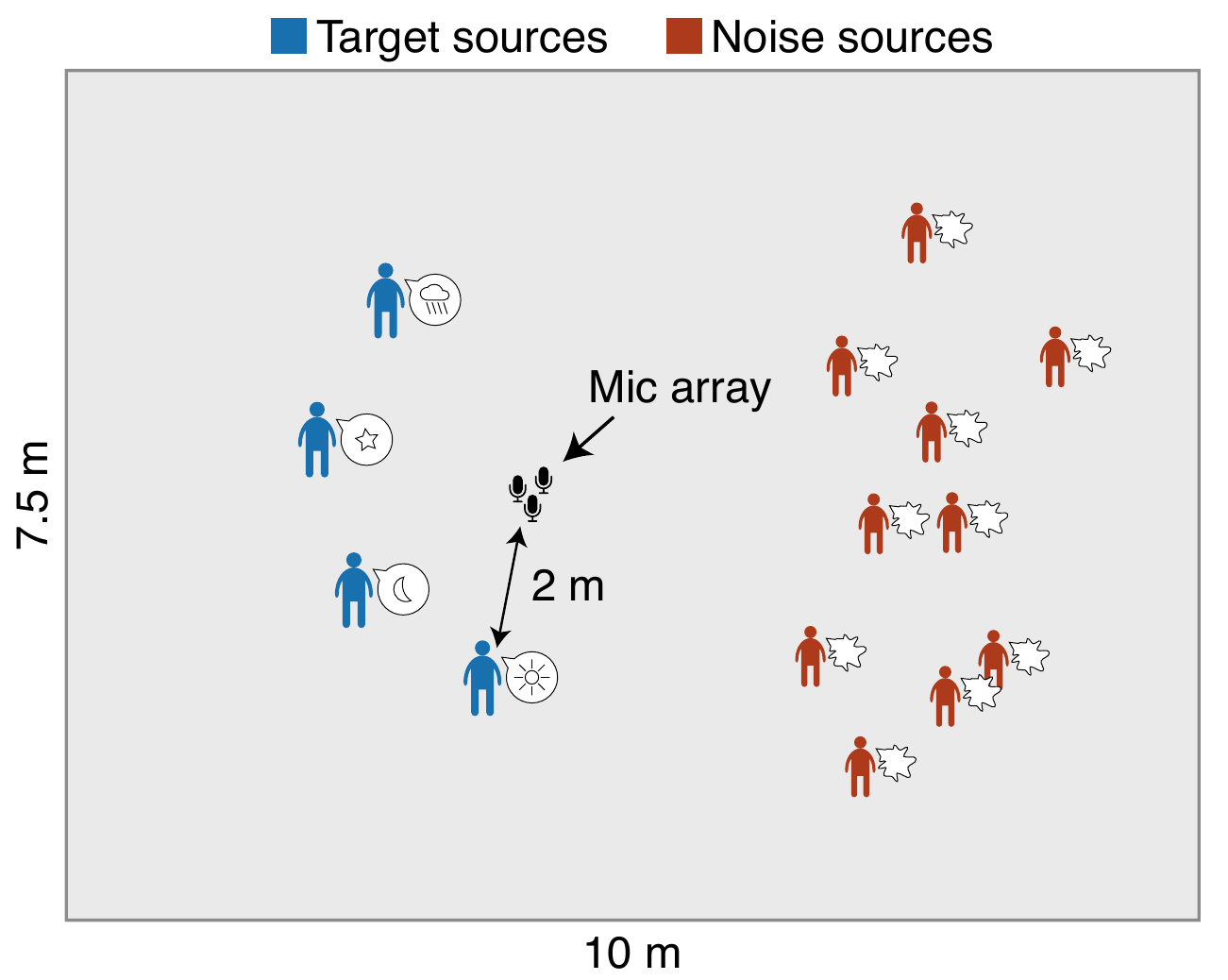}
\end{center}
\caption{Setup of the simulated experiment.}
\flabel{experiment_setup}
\end{figure}

\section{Performance Evaluation}
\seclabel{perfeval}

In this section, the separation and runtime performances of the proposed and conventional algorithms are compared.

\begin{figure*}
\begin{center}
  \includegraphics[width=\linewidth]{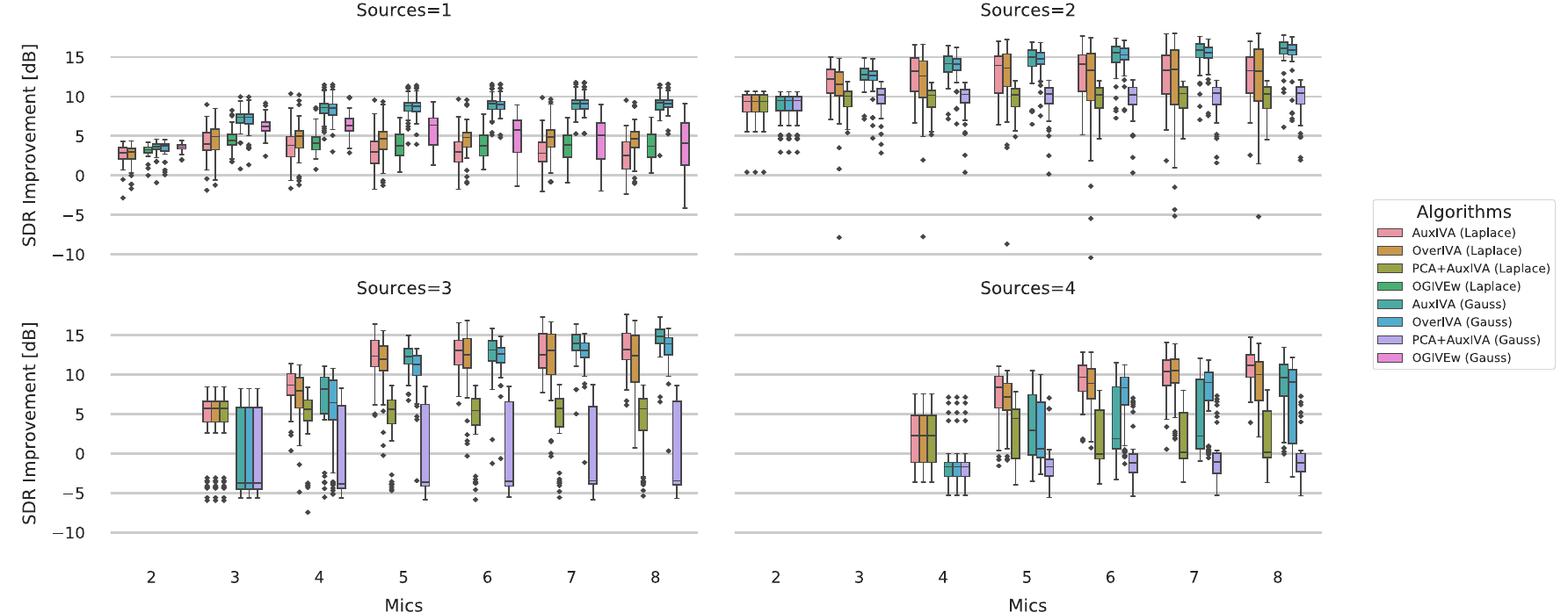}
\end{center}
\caption{
  Box-plots of signal-to-distortion ratio (SDR, top row) improvements between mixture and separated signals.
  Dots represent outliers.
  The number of sources increases from 1 to 4 left to right and top to bottom.
  The number of microphones increases from 2 to 8 on the horizontal axis.
}
\flabel{sep_perf}
\end{figure*}

\subsection{Setup}

We simulate a \SI{10}{\meter}$\times$\SI{7.5}{\meter}$\times$\SI{3}{\meter} room with reverberation time of \SI{300}{\milli\second} using the image source method~\cite{Allen:1979cn} implemented in the \texttt{pyroomacoustics} Python package~\cite{scheibler2018pyroomacoustics}.
We place a half-circular microphone array of radius \SI{4}{\centi\meter} at $[4.1, 3.76, 1.2]$.
The number of microphones is varied from 2 to 8.
Between 2 and 4 target sources are placed equispaced on an arc of \SI{120}{\degree} of radius \SI{2}{\meter} centered at the microphone array and at a height of \SI{1.2}{\meter}.
Diffuse noise is created by 10 additional sources on the opposite side of room.
This setup, illustrated in \ffref{experiment_setup}, is that of a few speakers holding a meeting in a noisy open office.

After simulating propagation, the variances of target sources are fixed to $\sigma_k^2 = 1$ (at an arbitrary reference microphone).
The signal-to-noise and signal-to-interference-and-noise ratios are defined as
\begin{equation}
  \mathsf{SNR} = \frac{\frac{1}{K} \sum_{k=1}^K \sigma_k^2}{\sigma_n^2},\quad \mathsf{SINR} = \frac{\sum_{k=1}^K \sigma^2_k}{Q \sigma_i^2 + \sigma_n^2},
\end{equation}
where $\sigma_i^2$ and $\sigma_n^2$ are the variances of the $Q$ interfering sources and uncorrelated white noise, respectively.
We set them so that $\mathsf{SNR}=60$ dB and $\mathsf{SINR}=10$ dB.
Speech samples of approximately \SI{20}{\second} are created by concatenating utterances from the CMU Sphinx database \cite{Kominek:2004vf}.
The experiment is repeated $50$ times for different attributions of speakers and speech samples to source locations.
The simulation is conducted at a sampling frequency of \SI{16}{\kilo\hertz}.
The STFT frame size is 4096 samples with half-overlap and uses a Hann window for analysis and matching synthesis window.
We compare OverIVA to three methods.
\begin{enumerate}
	\item \textbf{AuxIVA}: Full IVA with $M$ channels, followed by picking the $K$ strongest outputs.
	\item \textbf{PCA+AuxIVA}: Reduce the number of channels to $K$ via PCA, followed by IVA. This is only done when $K\geq2$.
  \item \textbf{OGIVEw}: For $K=1$, orthogonally constrained independent vector extraction (OGIVEw)~\cite{Koldovsky:fn}.
\end{enumerate}
We further compare the time-varying Gauss and Laplace versions of all these algorithms.
AuxIVA-based algorithms are run for 100 iterations.
OGIVEw is run for 4000 iterations with step size of 0.01.
The scale of the separated signals is restored by projecting back on the first microphone~\cite{Murata:2001gb}.

\begin{figure}
\begin{center}
  \includegraphics[scale=0.885]{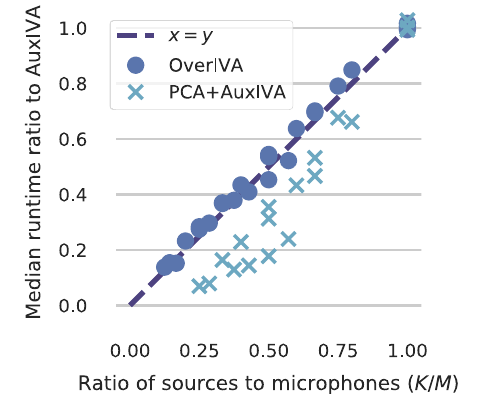}
  \includegraphics[scale=0.885]{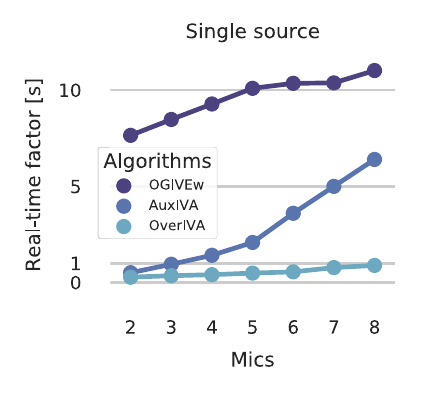}
\end{center}
\caption{Left, ratio of median runtimes of OverIVA/PCA+AuxIVA to full AuxIVA. Right, runtime per second of audio (i.e., real-time factor) for single source extraction.}
\flabel{runtime_perf}
\end{figure}

\subsection{Runtime Performance}

To verify the claim of \sref{comp}, we measured the runtime of 100 runs of each algorithm and compute the median.
As shown in \ffref{runtime_perf}, on the left, the ratio of the runtime of OverIVA to that of AuxIVA follows closely the predicted $K/M$.
Unsurprisingly, PCA+AuxIVA is much more computationally efficient since it only performs IVA on $K$ channels.
However, its separating performance falls short as discussed in the next section.

For a single source, as shown in \ffref{runtime_perf}, right, OverIVA is very fast and has real-time factor (RTF) less than one for up to 8 microphones (using 100 iterations).
Comparatively, AuxIVA has RTF less than one only up to 3 microphones.
We also find that our straightforward Python implementation of OGIVEw is not competitive.
Let us note that OGIVEw requires many gradient ascent iterations that might run faster in a compiled language such as C or C++.

\subsection{Separation Performance}

The separation performance of the algorithms is assessed in terms of signal-to-distortion ratio (SDR) as defined in \cite{Vincent:2006fz}.
These metrics are computed using the \texttt{mir\_eval} toolbox \cite{Raffel:2014uu}.
\ffref{sep_perf} shows box-plots of SDR improvements (with respect to the mixture signal).

We find that of all algorithms, OverIVA and AuxIVA perform best and similarly over all cases investigated.
It is interesting to notice a large gap between the determined case (where both algorithms are identical) and using one extra microphone.
Just the one extra input signal boosts SDR by 3 to 4~dB.
Adding further microphones consistently improves SDR, albeit at a slower pace.
In the single source extraction scenario (i.e., $K=1$), OverIVA turns out to be perfectly suitable and largely outperforms the state-of-the-art method OGIVEw.
When $K\geq 2$, the PCA+AuxIVA method falls short in terms of separation, with virtually no improvement when using more microphones.
This is likely due to the diffuse noise, since PCA is only optimal when the noise is uncorrelated across channels.
Finally, the difference between using Gauss or Laplace models seems consistent across algorithms.
For 1 and 2 sources, Gauss IVA performs better than Laplace IVA.
However, the trend reverses for 3 and 4 sources.
We conjecture that the Laplace AuxIVA might be more robust to mismatched initialization.
Using more microphones seems to make the gap in performance disappear.

\section{Conclusion}
\seclabel{conclusion}

We introduced OverIVA, a hyperparameter-free algorithm for blind source separation with more microphones than sources.
The algorithm applies the efficient updates from auxiliary function-based IVA while maintaining orthogonality between the signal and noise subspaces.
A parametrization of the demixing matrix that reduces the number of parameters to estimate is introduced to reduce complexity.
We show that using more microphones indeed increases, sometimes dramatically, performance, and that OverIVA solves the problem at a fraction of the cost of full IVA.
We also verify that the algorithm performs largely over the state-of-the-art in the so-called blind source extraction (single source) case.
Future work will focus on applying the algorithm to recorded data and assessing its performance for real-time implementation.

\bibliographystyle{IEEEtran}
\bibliography{refs}

\end{document}